\documentclass[natbib,epj]{svjour}

\usepackage{amsmath}
\usepackage{amssymb}
\usepackage{amsfonts}
\usepackage{graphicx}% Include figure files
\usepackage{dcolumn}% Align table columns on decimal point
\usepackage{bm}% bold math
\usepackage{color}% color

\begin{document}

\title{Quantifying bid-ask spreads in the Chinese stock market using limit-order book data}%

\subtitle{Intraday pattern, probability distribution, long memory, and multifractal nature}%

\author{Gao-Feng Gu\inst{1,2} \and Wei Chen\inst{3} \and Wei-Xing Zhou\inst{1,2,4,}\thanks{e-mail:
wxzhou@ecust.edu.cn}}

\institute{School of Business, East China University of Science and
Technology, Shanghai 200237, China \and School of Science, East
China University of Science and Technology, Shanghai 200237, China
\and Shenzhen Stock Exchange, 5045 Shennan East Road, Shenzhen
518010, China \and Research Center of Systems Engineering, East
China University of Science and Technology, Shanghai 200237, China}

\date{Received: January 5, 2007 / Revised version:  \today}

\abstract{The statistical properties of the bid-ask spread of a
frequently traded Chinese stock listed on the Shenzhen Stock
Exchange are investigated using the limit-order book data. Three
different definitions of spread are considered based on the time
right before transactions, the time whenever the highest buying
price or the lowest selling price changes, and a fixed time
interval. The results are qualitatively similar no matter linear
prices or logarithmic prices are used. The average spread exhibits
evident intraday patterns consisting of a big L-shape in morning
transactions and a small L-shape in the afternoon. The distributions
of the spread with different definitions decay as power laws. The
tail exponents of spreads at transaction level are well within the
interval $(2,3)$ and that of average spreads are well in line with
the inverse cubic law for different time intervals. Based on the
detrended fluctuation analysis, we found the evidence of long memory
in the bid-ask spread time series for all three definitions, even
after the removal of the intraday pattern. Using the classical
box-counting approach for multifractal analysis, we show that the
time series of bid-ask spread does not possess multifractal nature.
}

\maketitle

\section{Introduction}

The continuous double auction (CDA) is a dominant market mechanism
used to store and match orders and to facilitate trading in most
modern equity markets
\cite{Smith-Farmer-Gillemot-Krishnamurthy-2003-QF}. In most of the
order driven markets, there are two kinds of basic orders, called
market orders and limit orders. A market order is submitted to buy
or sell a number of shares at the market quote which results in an
immediate transaction, while a limit order is placed to buy (or
sell) a number of shares below (or above) a given price. All the
limit orders that fail to result in an immediate transaction are
stored in a queue called {\em{limit-order book}}. Buy limit orders
are called bids while sell limit orders are called asks or offers.
{\em{Best bid}} price $b(t)$ and {\em{best ask}} (or {\em{best
offer}}) price $a(t)$ are the highest buying price and the lowest
selling price at any time $t$ in the limit-order book. The best bid
(or ask) is called the {\em{same best}} for buy (or sell) orders,
while the best ask (or bid) is called the {\em{opposite best}} for
buy (or sell) orders. A limit order causes an immediate transaction
if the associated limit price penetrates the opposite best price.
Such kind of limit orders are called {\em{marketable limit orders}}
or {\em{effective market orders}} and other limit orders are termed
{\em{effective limit orders}}. In the Chinese stock market, only
limit orders were permitted in the placement of orders before July
1, 2006.

It is a dynamic process concerning the limit-order book. Effective
limit orders accumulate in the book while effective market orders
cause transactions and remove the limit orders according to their
price and the time they arrive. Effective limit orders can also be
removed by cancelation for a variety of reasons. Unveiling the
dynamics of order placement and cancelation will deepen our
understanding of the microscopic mechanism of price formation and
allow us to reproduce remarkably many key features of common stocks
such as the probability distribution of returns
\cite{Zovko-Farmer-2002-QF,Gabaix-Gopikrishnan-Plerou-Stanley-2003-Nature,Bouchaud-Gefen-Potters-Wyart-2004-QF,Plerou-Gopikrishnan-Gabaix-Stanley-2004-QF,Farmer-Patelli-Zovko-2005-PNAS,Malevergne-Pisarenko-Sornette-2005-QF,Bouchaud-Kockelkoren-Potters-2006-QF,Mike-Farmer-2007-JEDC}.

The difference between best-ask price and best-bid price,
$s(t)=a(t)-b(t)$, is the bid-ask spread. Numerous work has been
carried out to explore the different components of the bid-ask
spread \cite{Stoll-1989-JF,Huang-Stoll-1997-RFS}. On the other hand,
there are several groups studying the statistical properties of the
bid-ask spread time series for different stock markets. Farmer
{\em{et al.}} reported that the bid-ask spread defined by
$\ln[a(t)]-\ln[b(t)]$ on the London Stock Exchange follows power-law
distribution in the tail
\begin{equation}
P(>s) \sim s^{-\zeta}~,
\end{equation}
where the exponent $\zeta=3.03\pm0.41$ ranging from 2.4 to 3.9
\cite{Farmer-Gillemot-Lillo-Mike-Sen-2004-QF,Mike-Farmer-2007-JEDC},
which is well consistent with the inverse cubic law
\cite{Gopikrishnan-Meyer-Amaral-Stanley-1998-EPJB,Gabaix-Gopikrishnan-Plerou-Stanley-2003-PA,Gabaix-Gopikrishnan-Plerou-Stanley-2003-Nature}.
In addition, Mike and Farmer found that the spread possesses long
memory with the Hurst index being $0.75<H<0.85$
\cite{Mike-Farmer-2007-JEDC}. Plerou {\em{et al.}} adopted the 116
most frequently traded stocks on the New York Stock Exchange over
the two-year period 1994-1995 to investigate the coarse-grained
bid-ask spread over a time interval $\Delta{t}$ and found that the
tail distribution decays as a power law with a mean tail exponent of
$\zeta=3.0\pm0.1$ and the spread after removing the intraday pattern
exhibits long memory with $H=0.73\pm0.01$
\cite{Plerou-Gopikrishnan-Stanley-2005-PRE}. Qualitatively similar
results were found by Cajueiro and Tabak in the Brazilian equity
market where the mean tail exponent is $\zeta=2.18$ ranging from
1.18 to 2.97 and the Hurst index is $H=0.68\pm0.08$ varying from
0.52 to 0.89 \cite{Cajueiro-Tabak-2007-PA}.

Due to the fast development of the economy of China and the
increasing huge capitalization of its stock market, more concerns
are attracted to study the emerging Chinese stock market. In order
to reduce the market risks and speculation actions, the Chinese
stock market adopts $t+1$ trading system, which does not allow
traders to sell and buy stocks on the same day, and no market orders
were permitted until July 1, 2006, which may however consume the
liquidity of the market and cause the spread to show different
properties when compared to other stock markets. In this work, we
investigated the probability distribution, long memory, and presence
of multifractal nature of the bid-ask spread using limit-order book
data on the Shenzhen Stock Exchange (SSE) in China.

The rest of this paper is organized as follows. In
Sec.~\ref{s1:database}, we describe in brief the trading rules of
the Shenzhen Stock Exchange and the database we adopt. Section
\ref{s1:definition} introduces three definitions of the bid-ask
spread and investigates the intraday pattern in the spread. The
cumulative distributions of the spreads for different definitions
are discussed in Sec.~\ref{s1:PDF}. We show in Sec.~\ref{s1:memory}
the long memory of the spread based on the detrended fluctuation
analysis (DFA) quantified by the estimate of the Hurst index. In
Sec.~\ref{s1:MFA}, we perform multifractal analysis on the bid-ask
spread time series. The last section concludes.

\section{SSE trading rules and the data set}
\label{s1:database}

Our analysis is based on the limit-order book data of a liquid stock
listed on the Shenzhen Stock Exchange. SSE was established on
December 1, 1990 and has been in operation since July 3, 1991. The
securities such as stocks, closed funds, warrants and Lofs can be
traded on the Exchange. The Exchange is open for trading from Monday
to Friday except the public holidays and other dates as announced by
the China Securities Regulatory Commission. With respect to
securities auction, opening call auction is held between 9:15 and
9:25 on each trading day, followed by continuous trading from 9:30
to 11:30 and 13:00 to 15:00. The Exchange trading system is closed
to orders cancelation during 9:20 to 9:25 and 14:57 to 15:00 of each
trading day. Outside these opening hours, unexecuted orders will be
removed by the system. During 9:25 to 9:30 of each trading day, the
Exchange is open to orders routing from members, but does not
process orders or process cancelation of orders.

Auction trading of securities is conducted either as a call auction
or a continuous auction. The term ``call auction'' (from 9:15 to
9:25) refers to the process of one-time centralized matching of buy
and sell orders accepted during a specified period in which the
single execution price is determined according to the following
three principles: (i) the price that generates the greatest trading
volume; (ii) the price that allows all the buy orders with higher
bid price and all the sell orders with lower offer price to be
executed; and (iii) the price that allows either buy orders or sell
orders to have all the orders identical to such price to be
executed.

The term ``continuous auction'' (from 9:25 to 11:30 and from 13:00
to 15:00) refers to the process of continuous matching of buy and
sell orders on a one-by-one basis and the execution price in a
continuous trading is determined according to the following
principles: (i) when the best ask price equals to the best bid
price, the deal is concluded at such a price; (ii) when the buying
price is higher than the best ask price currently available in the
central order book, the deal is concluded at the best ask price; and
(iii) when the selling price is lower than the best bid price
currently available in the central order book, the deal is executed
at the best bid price. The orders which are not executed during the
opening call auction automatically enter the continuous auction.

The tick size of the quotation price of an order for A
shares\footnote{{\textbf{A shares}} are common stocks issued by
mainland Chinese companies, subscribed and traded in Chinese RMB,
listed in mainland Chinese stock exchanges, bought and sold by
Chinese nationals. A-share market was launched in 1990.} is RMB 0.01
and that for B shares\footnote{{\textbf{B shares}} are issued by
mainland Chinese companies, traded in foreign currencies and listed
in mainland Chinese stock exchanges. B shares carry a face value
denominated in Renminbi. The B Share Market was launched in 1992 and
was restricted to foreign investors before February 19, 2001. B
share market has been opened to Chinese investors since February 19,
2001.} is HKD 0.01. Orders are matched and executed based on the
principle of {\em{price-time priority}} which means priority is
given to a higher buy order over a lower buy order and a lower sell
order is prioritized over a higher sell order; The order sequence
which is arranged according to the time when the Exchange trading
system receives the orders determines the priority of trading for
the orders with the same prices.

We studied the data from the limit-order book of the stock SZ000001
(Shenzhen Development Bank Co., LTD) in the whole year of 2003. The
limit-order book recorded high-frequency data whose time stamps are
accurate to 0.01 second. The size of the data set is $3,925,832$,
including $12,965$ invalid orders, $122,034$ order submissions and
cancelations in the opening call auction, $47,576$ order submissions
and cancelations during the cooling period (9:25-9:30), and
$3,743,257$ valid events during the continuous auction. In
continuous auction, there are $317,015$ cancelations of buy orders
and $274,929$ cancelations of sell orders, $889,700$ effective
market orders, and $2,261,613$ effective limit orders. Table
\ref{Tb:1} shows a segment taken from the limit-order book recorded
on 2003/07/09. The seven columns stand for order size, limit price,
time, best bid, best ask, transaction volume, and buy-sell
identifier (which identifies whether a record is a buy order, sell
order, or a cancelation). For a cancelation record, the limit price
is set to be zero.

\begin{table}[htp]
\caption{A segment of the limit-order book}\label{Tb:1}
\begin{center}
\begin{tabular}{rrrrrrrrrrrrr}
  \hline
 1400 &0    &9390015&11.33&11.34&0&31\\
 1000 &11.48&9390016&11.33&11.34&0&29\\
 400  &11.65&9390311&11.33&11.34&0&29\\
 400  &0    &9390317&11.33&11.34&0&30\\
 1000 &11.33&9390365&11.33&11.34&0&26\\
 6000 &11.33&9390408&11.33&11.34&6000&23\\
 \hline
\end{tabular}
\end{center}
%\vspace*{5mm}
\end{table}

\section{Defining bid-ask spread}
\label{s1:definition}

The literature concerning the bid-ask spread gives different
definitions
\cite{Farmer-Patelli-Zovko-2005-PNAS,Mike-Farmer-2007-JEDC,Stoll-1989-JF,Huang-Stoll-1997-RFS,Farmer-Gillemot-Lillo-Mike-Sen-2004-QF,Plerou-Gopikrishnan-Stanley-2005-PRE,Cajueiro-Tabak-2007-PA,Roll-1984-JF,Daniels-Farmer-Gillemot-Iori-Smith-2003-PRL,Wyart-Bouchaud-Kockelkoren-Potters-Vettorazzo-2006}.
In this section, we discuss three definitions according to sampling
time when best bid prices and best ask prices are selected to define
the spread. Some definitions are based on the transaction time,
while the others are based on the physical time. The latter scheme
is actually a coarse-graining of the data within a given time
interval.

\subsection{Definition I}

The first definition of the bid-ask spread used in this work is the
absolute or relative difference between the best ask price and the
best bid price right before the transaction, that is,
\begin{subequations}\label{Eq:s12}
\begin{equation}
  s(t) = a(t) - b(t)
\end{equation}
for absolute difference or
\begin{equation}
  s(t)=\log_{10} [a(t)] -\log_{10} [b(t)]
\end{equation}
\end{subequations}
for relative difference. This was used to analyze the stocks on the
London Stock Exchange
\cite{Farmer-Gillemot-Lillo-Mike-Sen-2004-QF,Mike-Farmer-2007-JEDC}.
The size of the spread time series is $895,606$.

\subsection{Definition II}

The best ask price or the best bid price may change due to the
removal of all shares at the best price induced by an effective
market order, or the placement of an limit order inside the spread,
or the cancelation of all limit orders at the best bid/ask price.
Hence the bid-ask spread does not always change when a transaction
occurs, and it nevertheless changes without transaction. This
suggests to introduce an alternative definition of the spread which
considers the absolute or relative difference between the best bid
price and the best ask price whensoever it changes. The expressions
of definition II are the same as those in Eq.~(\ref{Eq:s12}) except
that they have different definitions for the time $t$. The size of
the spread time series is $142,913$.

\subsection{Definition III}

Obviously, the time in the first two definitions are on the basis of
``event''. An alternative definition considers the average bid-ask
spread over a time interval  $\Delta{t}$
\cite{Mcinish-Wood-1992-JF}. In this definition, the bid-ask spread
is the average difference between the best ask the best bid when
transactions occur over a fixed time interval
\cite{Plerou-Gopikrishnan-Stanley-2005-PRE}:
\begin{equation}
s(t) = \frac{1}{N}\sum_{i=1}^N{s_i(t)}~,~~~s_i(t) = a_i(t) -
b_i(t)~,
 \label{Eq:s3}
\end{equation}
where $a_i(t)$ and $b_i(t)$ are the best ask and bid prices in the
time interval $(t-{\Delta} t,t]$, and $N$ is the total number of
transaction in the interval and is a function of $t$ and
$\Delta{t}$. We use ${\Delta} t= 1$, $2$, $3$, $4$, and $5$
minute(s) to calculate the average spreads.

\subsection{Intraday pattern}

In most modern financial markets, the intraday pattern exists
extensively in many financial variables
\cite{Wood-McInish-Ord-1985-JF,Harris-1986-JFE,Admati-Pfleiderer-1988-RFS},
including the bid-ask spread \cite{Mcinish-Wood-1992-JF}. The
periodic pattern has significance impact on the detection of long
memory in time series
\cite{Hu-Ivanov-Chen-Carpena-Stanley-2001-PRE}. To the best of our
knowledge, the investigation of the presence of intraday pattern in
the spreads of Chinese stocks is lack.

\begin{figure}[htb]
\begin{center}
\includegraphics[width=8cm]{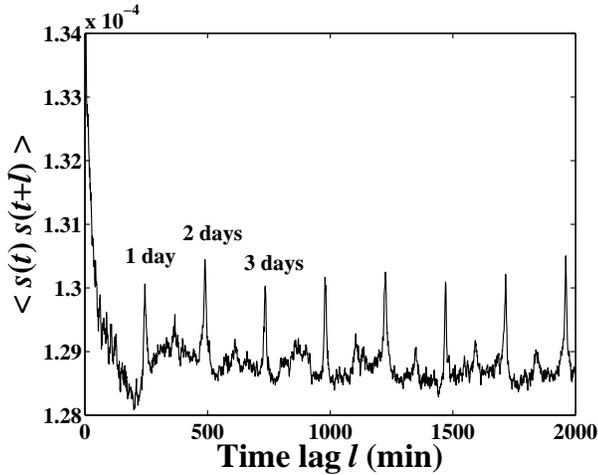}
\caption{Autocorrelation function $\langle{s(t)s(t+{\ell})}\rangle$
of the average bid-ask spread calculated from definition III with
the time interval $\Delta{t}=1$ min. Note that one trading day
contains 245 trading minutes in the Chinese stock market.}
\label{Fig:autocorrelation}
\end{center}
\end{figure}

Figure~\ref{Fig:autocorrelation} shows the autocorrelation function
$\langle{s(t)s(t+{\ell})}\rangle$ as a function of the time lag
$\ell$ for the average bid-ask spread calculated from definition III
with linear best bids and asks. We note that the results are very
similar when logarithmic prices are adopted in the definition. We
see that there are spikes evenly spaced along multiples of 245 min,
which is exactly the time span of one trading day. What is
interesting is that Fig.~\ref{Fig:autocorrelation} indicates that
the average spread also possesses half-day periodicity.

\begin{figure}[htb]
\begin{center}
\includegraphics[width=8cm]{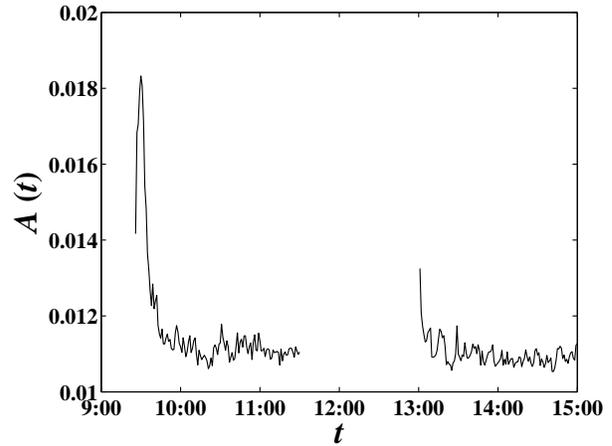}
\caption{Intraday pattern in the bid-ask spread with $\Delta{t}=1$
min. The spread reaches its maximum at the end of the cooling period
at 9:30.} \label{Fig:intraday}
\end{center}
\end{figure}

In order to quantify the intraday pattern, we introduce a variable
$A(t)$, which is defined as the average bid-ask spread at time $t$
for all the trading days, that is,
\begin{equation}
A(t) = \frac{1}{M}\sum^{M}_{j=1}{s^j(t)}~,
 \label{Eq:intraday}
\end{equation}
where $M$ is the number of trading days in the data set and $s^j(t)$
is the bid-ask spread at time $t$ of day $j$. The spread $S(t)$
after removing the intraday pattern reads
\cite{Plerou-Gopikrishnan-Stanley-2005-PRE}
\begin{equation}
S(t) = s(t) / A(t)~.
 \label{Eq:S}
\end{equation}
Figure~{\ref{Fig:intraday}} illustrates the intraday pattern of the
bid-ask spread with ${\Delta} t = 1$ minute. The overall plot shows
an evident L-shaped pattern, which is consistent with the one-day
periodicity shown in the autocorrelation function in
Fig.~\ref{Fig:autocorrelation}. After the opening call auction, the
spread $A(t)$ widens rapidly and reaches its maximum 0.0183 at the
end of the cooling auction (9:30)\footnote{In 2003, there were three
best prices at each side disposed in 9:25 and remained unchanged
during the cooling period. Hence, the spreads shown in
Figure~{\ref{Fig:intraday}} during this period are virtually
genrated according to the trading mechanism.}. Then it decreases
sharply in fifteen minutes and becomes flat at a level of
$0.0112\pm0.0008$ afterwards till 11:30. At the begin of continuous
auction in the afternoon, $A(t)$ abruptly rises to 0.0133 and drops
down to a stable level within about ten minutes which maintains
until the closing time 15:00. Therefore, there are two L-shaped
patterns each day, which suggests that the wide spread is closely
related to the opening of the market. The intraday pattern makes no
difference when we use ${\Delta} t = 2$, $3$, $4$, and $5$ minutes.

\section{Probability distribution}
\label{s1:PDF}

\begin{figure}[htb]
\begin{center}
\includegraphics[width=8cm]{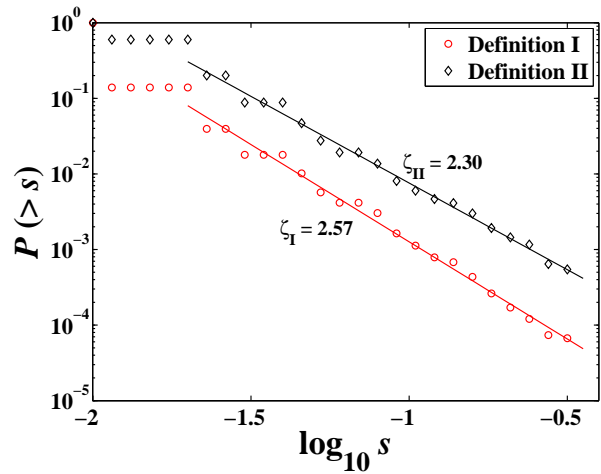}
\caption{Empirical complementary cumulative distribution of the
spreads calculated from definitions I and II using linear prices.}
\label{Fig:s12:cdf}
\end{center}
\end{figure}

The cumulative distributions of the bid-ask spread of stocks in
different stock markets decay as power laws with the tail exponent
close to 3 for the major western markets
\cite{Farmer-Gillemot-Lillo-Mike-Sen-2004-QF,Plerou-Gopikrishnan-Stanley-2005-PRE,Mike-Farmer-2007-JEDC}
and much smaller and more heterogeneous in an emerging market
\cite{Cajueiro-Tabak-2007-PA}. Similar behavior is found in the
Chinese stock market. Figure~\ref{Fig:s12:cdf} presents the
complementary cumulative distribution $P(\geqslant{s})$ of the
spreads using definition I and II, where linear prices are used.
Since the minimum spread equals to the tick size 0.01, the abscissa
is no less than -2 in double logarithmic coordinates and
$P(\geqslant0.01)=1$ for both definitions. The proportion of
$s=0.01$ in the first definition is much greater than in the second
definition such that the $P(\geqslant{s})$ for the second defintion
drops abruptly for small spreads $s$. The two distributions decay as
power laws with exponents $\zeta_{\rm{I}} = 2.57 \pm 0.06$ for
definition I and $\zeta_{\rm{II}} = 2.30 \pm 0.05$ for definition
II. When logarithmic prices are utilized, the spreads also follow
power-law tail distributions with $\zeta_{\rm{I}} = 2.67 \pm 0.03$
for definition I and $\zeta_{\rm{II}} = 2.42 \pm 0.04$ for
definition II. Not much difference in the corresponding tail
exponents $\zeta_{\rm{I}}$ and $\zeta_{\rm{II}}$ was found for
logarithmic and linear prices.

\begin{figure}[htb]
\begin{center}
\includegraphics[width=8cm]{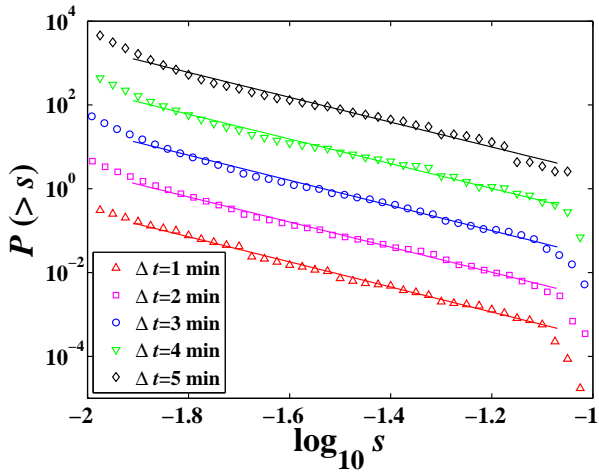}
\caption{Empirical complementary cumulative distributions of the
average spreads calculated from definition III with time intervals
${\Delta} t= 1$, $2$, $3$, $4$, and $5$ min using linear prices. The
markers represent the real data and the solid lines are the best
fits in the scaling ranges. The curves with $\Delta{t}>1$ has been
translated vertically for clarity.} \label{Fig:s3:cdf}
\end{center}
\end{figure}

Figure~\ref{Fig:s3:cdf} illustrates the complementary cumulative
distributions of the average spreads over time interval $\Delta{t}=
1$, $2$, $3$, $4$, and $5$ minute(s) calculated from definition III
with linear prices. The average spreads have power-law tails with
the exponents equal to $\zeta_{\rm{III},1} = 2.99 \pm 0.04$,
$\zeta_{\rm{III},2} = 3.00 \pm 0.04$, $\zeta_{\rm{III},3} = 3.00 \pm
0.05$, $\zeta_{\rm{III},4} = 2.95 \pm 0.05$, and $\zeta_{\rm{III},5}
= 2.97 \pm 0.06$. Similarly, for logarithmic prices, we find similar
power-law tail distributions with $\zeta_{\rm{III},1} = 3.07 \pm
0.06$, $\zeta_{\rm{III},2} = 2.95 \pm 0.05$, $\zeta_{\rm{III},3} =
3.00 \pm 0.04$, $\zeta_{\rm{III},4}= 2.97 \pm 0.07$, and
$\zeta_{\rm{III},5} = 2.98 \pm 0.07$. We find that all the tail
exponents $\zeta_{{\rm{III}},\Delta{t}}$ for both linear and
logarithmic prices are very close to three and are independent to
the time interval $\delta{t}$, showing a nice inverse cubic law.
This is well in agreement with the results in the NYSE case for
$\Delta{t}=15$, $30$, and $60$ min
\cite{Plerou-Gopikrishnan-Stanley-2005-PRE}.

There are also significant discrepancies. Comparing the cumulative
distributions in Fig.~\ref{Fig:s3:cdf} and that on the NYSE
\cite{Plerou-Gopikrishnan-Stanley-2005-PRE}, significant differences
are observed. The distribution of the spreads on the SSE decays much
faster than that on the NYSE for small spreads. In other words, the
proportion of small spreads is much larger on China's SSE. Possible
causes include the absence of market orders, no short positions, the
maximum percentage of fluctuation (10\%) in each day, and the $t+1$
trading mechanism in the Chinese stock markets on the one hand and
the hybrid trading system containing both specialists and
limit-order traders in the NYSE on the other hand. The exact cause
is not clear for the time being, which can however be tested when
new data are available after the introduction of market orders in
July 1, 2006. Moreover, the PDF's in SSE drop abruptly after the
power-law parts for the largest spreads, which is not observed in
the NYSE case \cite{Plerou-Gopikrishnan-Stanley-2005-PRE}.

\section{Long memory}
\label{s1:memory}

Another important issue about financial time series is the presence
of long memory, which can be characterized by its Hurst index $H$.
If $H$ is significantly larger than $0.5$ the time series is viewed
to possess long memory. Long memory can be defined equivalently
through autocorrelation function $C(\ell) \sim \ell^{-{\gamma}}$ and
the power spectrum $p(\omega)\sim\omega^{-\eta}$, where the
autocorrelation exponent ${\gamma}$ is related to the Hurst index
$H$ by $\gamma=2-2H$
\cite{Kantelhardt-Bunde-Rego-Havlin-Bunde-2001-PA,Maraun-Rust-Timmer-2004-NPG},
and the power spectrum exponent $\eta$ is given by $\eta=2H-1$
\cite{Talkner-Weber-2000-PRE,Heneghan-McDarby-2000-PRE}.

There are many methods proposed for estimating the Hurst index such
as the rescaled range analysis (RSA)
\cite{Hurst-1951-TASCE,Mandelbrot-Ness-1968-SIAMR,Mandelbrot-Wallis-1969a-WRR,Mandelbrot-Wallis-1969b-WRR,Mandelbrot-Wallis-1969c-WRR,Mandelbrot-Wallis-1969d-WRR},
fluctuation analysis (FA)
\cite{Peng-Buldyrev-Goldberger-Havlin-Sciortino-Simons-Stanley-1992-Nature},
detrended fluctuation analysis (DFA)
\cite{Peng-Buldyrev-Havlin-Simons-Stanley-Goldberger-1994-PRE,Hu-Ivanov-Chen-Carpena-Stanley-2001-PRE,Kantelhardt-Bunde-Rego-Havlin-Bunde-2001-PA},
wavelet transform module maxima (WTMM) method
\cite{Holschneider-1988-JSP,Muzy-Bacry-Arneodo-1991-PRL,Muzy-Bacry-Arneodo-1993-JSP,Muzy-Bacry-Arneodo-1993-PRE,Muzy-Bacry-Arneodo-1994-IJBC},
detrended moving average (DMA)
\cite{Alessio-Carbone-Castelli-Frappietro-2002-EPJB,Carbone-Castelli-Stanley-2004-PA,Carbone-Castelli-Stanley-2004-PRE,Alvarez-Ramirez-Rodriguez-Echeverria-2005-PA,Xu-Ivanov-Hu-Chen-Carbone-Stanley-2005-PRE},
to list a few. We adopt the detrended fluctuation analysis.

The method of detrended fluctuation analysis is widely used for its
easy implementation and robust estimation even for a short time
series
\cite{Taqqu-Teverovsky-Willinger-1995-Fractals,Montanari-Taqqu-Teverovsky-1999-MCM,Heneghan-McDarby-2000-PRE,Audit-Bacry-Muzy-Arneodo-2002-IEEEtit}.
The idea of DFA was invented originally to investigate the
long-range dependence in coding and noncoding DNA nucleotides
sequence\cite{Peng-Buldyrev-Havlin-Simons-Stanley-Goldberger-1994-PRE}
and then applied to various fields including finance. The method of
DFA consists of the following steps.

Step 1: Consider a time series $x(t)$, $t=1,2,\cdots,N$. We first
construct the cumulative sum
\begin{equation}
u(t) = \sum_{i = 1}^{t}{x(i)}, ~~t=1,2,\cdots,N~.
 \label{Eq:DFA_u}
\end{equation}

Step 2: Divide the series $u(t)$ into $N_\ell$ disjoint segments
with the same length $\ell$, where $N_\ell = [N/\ell]$. Each segment
can be denoted as $u_v$ such that $u_v(i) = u(l + i)$ for
$1\leqslant{i}\leqslant{\ell}$, and $l = (v - 1)\ell$. The trend of
$u_v$ in each segment can be determined by fitting it with a linear
polynomial function $\widetilde{u}_v$. Quadratic, cubic or higher
order polynomials can also be used in the fitting procedure while
the simplest function could be linear. In this work, we adopted the
linear polynomial function to represent the trend in each segment
with the form:
\begin{equation}
\widetilde{u}_v(i) = ai+b~,
 \label{Eq:DFA_wu}
\end{equation}
where $a$ and $b$ are free parameters to be determined by the least
squares fitting method and $1\leqslant{i}\leqslant\ell$.

Step 3: We can then obtain the residual matrix $\epsilon_{v}$ in
each segment through:
\begin{equation}
\epsilon_{v}(i)=u_{v}(i)-\widetilde{u}_{v}(i)~,
\end{equation}
where $1\leqslant{i}\leqslant{\ell}$. The detrended fluctuation
function $F(v,\ell)$ of the each segment is defined via the sample
variance of the residual matrix $\epsilon_{v}$ as follows:
\begin{equation}
F^2(v,\ell) = \frac{1}{\ell}\sum_{i = 1}^{\ell}[\epsilon_{v}(i)]^2~.
 \label{Eq:DFA_F1}
\end{equation}
Note that the mean of the residual is zero due to the detrending
procedure.

Step 4: Calculate the overall detrended fluctuation function
$F(\ell)$, that is,
\begin{equation}
F^2(\ell) = \frac{1}{N_\ell}\sum_{v = 1}^{N_\ell}F^2(v,\ell)~.
 \label{Eq:DFA_F2}
\end{equation}

Step 5: Varying the value of $\ell$, we can determine the scaling
relation between the detrended fluctuation function $F(\ell)$ and
the size scale $\ell$, which reads
\begin{equation}
F(\ell) \sim \ell^{H}~,
 \label{Eq:DFA_H}
\end{equation}
where $H$ is the Hurst index of the time series
\cite{Taqqu-Teverovsky-Willinger-1995-Fractals,Kantelhardt-Bunde-Rego-Havlin-Bunde-2001-PA}.

\begin{figure}[htb]
\begin{center}
\includegraphics[width=8cm]{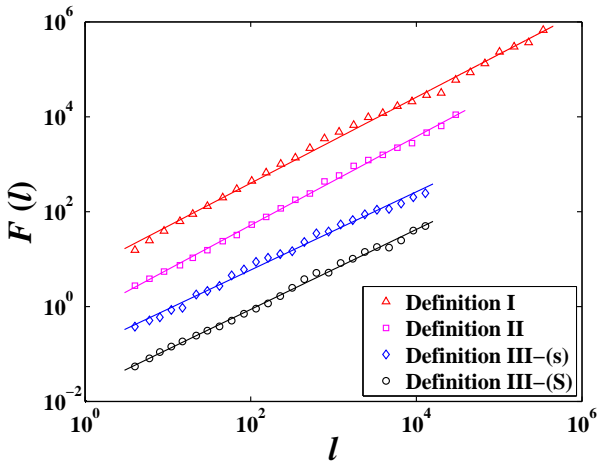}
\caption{Detrended fluctuation function $F(\ell)$ for the spreads
obtained from three definition with linear prices. The curves have
been shifted vertically for clarity.} \label{Fig:dfa}
\end{center}
\end{figure}

Figure~\ref{Fig:dfa} plots the detrended fluctuation function
$F(\ell)$ of the bid-ask spreads from different definitions using
linear prices. The bottom $F(\ell)$ curve is for the average spread
after removing the intraday pattern. All the curves show evident
power-law scaling with the Hurst indexes $H_{\rm{I}} = 0.91 \pm
0.01$ for definition I, $H_{\rm{II}} = 0.92 \pm 0.01$ for definition
II, $H_{\rm{III}} = 0.75 \pm 0.01$ for definition III, and
$H_{\rm{III}} = 0.77 \pm 0.01$ for definition without intraday
pattern, respectively. Quite similar results are obtain for
logarithmic prices where $H_{\rm{I}} = 0.89 \pm 0.01$ for definition
I, $H_{\rm{II}} = 0.91 \pm 0.01$ for definition II, $H_{\rm{III}} =
0.77 \pm 0.01$ for definition III, and $H_{\rm{III}} = 0.76 \pm
0.01$ for definition III without intraday pattern. The two Hurst
indexes for definitions I and II are higher than their counterparts
on the London Stock Exchange where ``even time'' is adopted
\cite{Mike-Farmer-2007-JEDC}. It is interesting to note that the
presence of intraday pattern does not introduce distinguishable
difference in the Hurst index and the two indexes for definition III
are also very close to those of average spreads in the Brazilian
stock market and on the New York Stock Exchange where real time is
used
\cite{Plerou-Gopikrishnan-Stanley-2005-PRE,Cajueiro-Tabak-2007-PA}.
Due to the large number of data used in the analysis, we argue that
the bid-ask spreads investigated exhibit significant long memory.

\section{Multifractal analysis}
\label{s1:MFA}

In this section, we investigate whether the time series of bid-ask
spread obtained from definition III possesses multifractal nature.
The classical box-counting algorithm for multifractal analysis is
utilized and described below
\cite{Halsey-Jensen-Kadanoff-Procaccia-Shraiman-1986-PRA}.

Consider the spread time series $S(t)$, $t=1,2,\cdots,N$. First, we
divide the series $S(t)$ into $N_\ell$ disjoint segments with the
same length $\ell$, where $N_\ell = [N/\ell]$. Each segment can be
denoted as $S_v$ such that $S_v(i) = S(l + i)$ for
$1\leqslant{i}\leqslant{\ell}$, and $l = (v - 1)\ell$. The sum of
$S_v$ in each segment is calculated as follows,
\begin{equation}
\Gamma(v,\ell) = \sum_{i = 1}^{\ell}{S_v(i)},
~~v=1,2,\cdots,N_\ell~.
 \label{Eq:MF_F}
\end{equation}
We can then calculate the $q$th order partition function $\Gamma(q;
\ell)$ as follows,
\begin{equation}
\Gamma(q; \ell) = \sum_{v = 1}^{N_\ell}[\Gamma(v,\ell)]^q~.
 \label{Eq:MF_Fq}
\end{equation}
Varying the value of $\ell$, we can determine the scaling relation
between the partition function $\Gamma(q; \ell)$ and the time scale
$\ell$, which reads
\begin{equation}
\Gamma(q; \ell) \sim \ell^{\tau(q)}~.
 \label{Eq:DFA_M_h}
\end{equation}

\begin{figure}[htb]
\begin{center}
\includegraphics[width=7cm]{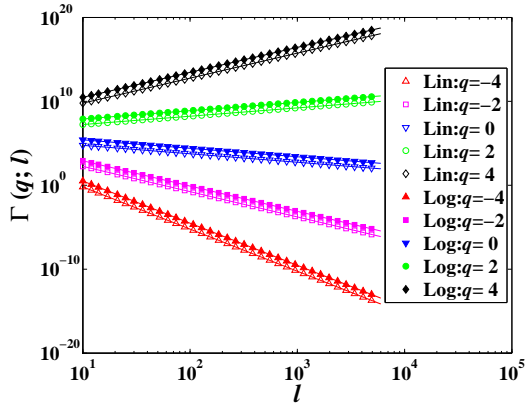}
\caption{Log-log plots of the partition function $\Gamma(q; \ell)$
of the bid-ask spreads calculated from definition III with the
intraday pattern removed for five different values of $q$. Both
linear and logarithmic prices are investigated (shown in the
legend). The markers stand for the results calculated from the real
data and the continuous lines are the best fits. The plots for $q =
-2, 0, 2$, and $4$ are shifted upwards for clarity.}
 \label{Fig:mffq}
\end{center}
\end{figure}

Figure~\ref{Fig:mffq} illustrates the power-law scaling dependence
of the partition function $\Gamma(q; \ell)$ of the bid-ask spreads
after removing the intraday pattern in definition III for different
values of $q$, where both linear prices and logarithmic prices are
investigated. The continuous lines are the best linear fits to the
data sets. The collapse of the data points on the linear lines
indicates evident power-law scaling between $\Gamma(q; \ell)$ and
$\ell$. The slopes $\tau(q)$ of the fitted lines are $\tau(-4) =
-5.02 \pm 0.01$, $\tau(-2) = -3.01 \pm 0.01$, $\tau(0) = -1.01 \pm
0.01$, $\tau(2) = 0.99 \pm 0.01$, and $\tau(4) = 2.98 \pm 0.01$ for
logarithmic prices and $\tau(-4) = -5.02 \pm 0.01$, $\tau(-2) =
-3.01 \pm 0.01$, $\tau(0) = -1.01 \pm 0.01$, $\tau(2) = 0.99 \pm
0.01$, and $\tau(4) = 2.98 \pm 0.01$ for linear prices. We notice a
nice relation $\tau(q)=q-1$.

Quantitatively similar results are obtained when the intraday
pattern is not removed. The scaling exponents are $\tau(-4) = -5.03
\pm 0.01$, $\tau(-2) = -3.01 \pm 0.01$, $\tau(0) = -1.01 \pm 0.01$,
$\tau(2) = 0.99 \pm 0.01$, and $\tau(4) = 2.96 \pm 0.01$ for
logarithmic prices and $\tau(-4) = -5.03 \pm 0.01$, $\tau(-2) =
-3.02 \pm 0.01$, $\tau(0) = -1.01 \pm 0.01$, $\tau(2) = 0.99 \pm
0.01$, and $\tau(4) = 2.97 \pm 0.01$ for linear prices. Again, we
observe that $\tau(q)=q-1$.

In the standard multifractal formalism based on partition function,
the multifractal nature is characterized by the scaling exponents
$\tau(q)$. It is easy to obtain the generalized dimensions $D_q=
{\tau}(q)/(q - 1)$
\cite{Grassberger-1983-PLA,Hentschel-Procaccia-1983-PD,Grassberger-Procaccia-1983-PD}
and the singularity strength function $\alpha(q)$, the multifractal
spectrum $f(\alpha)$ via the Legendre transform
\cite{Halsey-Jensen-Kadanoff-Procaccia-Shraiman-1986-PRA}:
$\alpha(q) = {\rm d}{\tau}(q)/{\rm d}q$ and $f(q) = q{\alpha} -
{\tau}(q)$.

\begin{figure}[htb]
\begin{center}
\includegraphics[width=8cm]{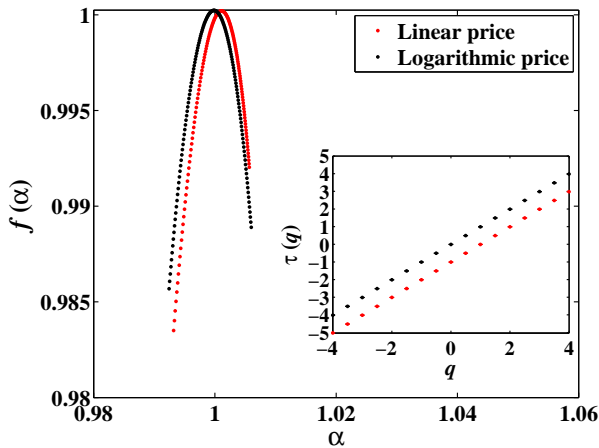}
\caption{Multifractal function $f({\alpha})$ of the spreads in
definition III with the intraday pattern removed. Inset: Scaling
exponents ${\tau}(q)$ of partition functions as a function of $q$.
For clarity, the $\tau(q)$ curve for logarithmic price is shifted
upwards by 1.}
 \label{Fig:falpha:tau}
\end{center}
\end{figure}

Figure~\ref{Fig:falpha:tau} shows the multifractal spectrum
$f(\alpha)$ and the scaling function $\tau(q)$ in the inset for
linear and logarithmic prices. One finds that the two $\tau(q)$
curves are linear and $\tau(q)=q-1$, which is the hallmark of the
presence of monofractality, not multifractality. The strength of the
multifractality can be characterized by the span of singularity
$\Delta\alpha=\alpha_{\max}-\alpha_{\min}$. If $\Delta\alpha$ is
close to zero, the measure is almost monofractal. The maximum and
minimum of $\alpha$ can be reached when $q\to\pm\infty$, which can
not be achieved in real applications. However, $\Delta\alpha$ can be
approximated with great precision with mediate values of $q$. The
small value of $\Delta\alpha$ shown in Fig.~\ref{Fig:falpha:tau}
indicates a very narrow spectrum of singularity. Indeed, one sees
that $f(\alpha)\approx1$ and $\alpha\approx1$ for all values of $q$.
We thus conclude that there is no multifractal nature in the bid-ask
spread investigated.

\section{Conclusions}
\label{s7:conclusion}

The bid-ask spread defined by the difference of the best ask price
and the best bid price is considered as the benchmark of the
transaction cost and a measure of the market liquidity. In this
paper, we have carried out empirical investigations on the
statistical properties of the bid-ask spread using the limit-order
book data of a stock SZ000001 (Shenzhen Development Bank Co., LTD)
traded on the Shenzhen Stock Exchange within the whole year of 2003.
Three different definitions of spread are considered based on event
time at transaction level and on fixed interval of real time.

The distributions of spreads at transaction level decay as power
laws with tail exponents well below 3. In contrast the average
spread in real time fulfils the inverse cubic law for different time
intervals $\Delta{t}= 1$, $2$, $3$, $4$, and $5$ min. We have
performed the detrended fluctuation analysis on the spread and found
that the spread time series exhibits evident long-memory, which is
in agreement with other stock markets. However, an analysis using
the classic textbook box-counting algorithm does not provide
evidence for the presence of multifractality in the spread time
series. To the best of our knowledge, this is the first time to
check the presence of multifractality in the spread.

Our analysis raises an intriguing open question that is not fully
addressed. We have found that the spread possesses a
well-established intraday pattern composed by a large L-shape and a
small L-shape separated by the noon closing of the Chinese stock
market. This feature will help to understand the cause of the wide
spread at the opening of the market, which deserves further
investigation.

\begin{acknowledgement}
We are grateful to Dr. Tao Wu for fruitful suggestions. This work
was partially supported by the National Natural Science Foundation
of China (Grant No. 70501011) and the Fok Ying Tong Education
Foundation (Grant No. 101086).
\end{acknowledgement}

\bibliographystyle{h-elsevier3}
%\bibliography{Bibliography}
\bibliography{E:/papers/Bibliography}
%\bibliography{D:/Bibliography/Bibliography}

\end{document}